# Unusual UUDD magnetic chain structure of the spin-1/2 tetragonally distorted spinel GeCu$_2$O$_4$

T. Zou[1,2], Y. -Q. Cai,[3] C. R. dela Cruz,[2] V. O. Garlea,[2] S. D. Mahanti,[1] J. -G. Cheng,[3] X. Ke[1]

[1]*Department of Physics and Astronomy, Michigan State University, East Lansing, MI 48824, USA*

[2]*Quantum Condensed Matter Division, Oak Ridge National Laboratory, Oak Ridge, TN 37831, USA*

[3]*Beijing National Laboratory for Condensed Matter Physics and Institute of Physics, Chinese Academy of Sciences, Beijing 100190, China*

GeCu$_2$O$_4$ exhibits a tetragonal spinel structure due to the strong Jahn-Teller distortion associated with Cu$^{2+}$ ions. We show that its magnetic structure can be described as slabs composed of a pair of layers with orthogonally oriented spin ½ Cu chains in the basal *ab* plane. The spins between the two layers within a slab are collinearly aligned while the spin directions of neighboring slabs are perpendicular to each other. Interestingly, we find that spins along each chain form an unusual up-up-down-down (UUDD) pattern, suggesting a non-negligible nearest-neighbor biquadratic exchange interaction in the effective classical spin Hamiltonian. We hypothesize that spin-orbit coupling and orbital mixing of Cu$^{2+}$ ions in this system is non-negligible, which calls for future calculations using perturbation theory with extended Hilbert (spin and orbital) space and calculations based on density functional theory including spin-orbit coupling and looking at the global stability of the UUDD state.

Spinel compounds, $AB_2O_4$, have been the subject of intense investigations due to the emergence of a rich variety of unconventional magnetic orderings [1], which arise from competing magnetic interactions and frustration associated with the geometry of transition-metal ions occupying the B sites of the spinel lattice that form a network of cornering-sharing tetrahedra (known as pyrochlore) [2, 3]. In addition, strong interplay between spin, lattice, and orbital degrees of freedom appears to play an important role in determining the magnetic ground state properties. It has been predicted that antiferromagnetically coupled Heisenberg spins on a pyrochlore lattice tend to form a spin liquid ground state [4]. However, the associated magnetic frustration can be readily relieved (partially) by the structural distortion (e.g., Jahn-Teller distortion) via spin-lattice coupling [5], single ion anisotropy [6], dipolar interaction [7], etc., which lifts both spin and orbital degeneracy and gives rise to long-range ordered spin solid.

In the past two decades, extensive studies on oxide spinels have focused on the role of the entanglement of spin and orbital degrees of freedom with the lattice structure in determining the physical properties of the so-called "2-3" spinels (containing divalent and trivalent cations, where $A^{2+}$ = Zn, Mg, Cd, Hg, Fe, Mn, Co and $B^{3+}$ = Ti, V, Cr, Fe, Co) [1]. Nevertheless, in general the cubic-to-tetragonal structural distortion in oxide spinels is relatively small and sometimes even challenging to be resolved, yet such a small change in the lattice is sufficient to lower the overall energy of the coupled lattice and magnetic system and leads to a magnetically ordered ground state [8-12]. As a result, most of the "2-3" spinels exhibit three-dimensional ordered spin arrangement, although a dimerized spin-singlet (spin liquid) ground state has been reported in $MgTi_2O_4$ where the structural distortion is found to be much larger [13].

In contrast, there are very limited studies on spinels with tetravalent A cation and divalent B cation, coined as "4-2" spinels. $GeTM_2O_4$ (TM = Fe, Co, Ni, Cu) are examples of such a class of compounds. Depending on the choice of TM ions on the B sites and the resultant outermost electronic configurations, these materials reveal interesting and diverse magnetic and structural properties. On the one hand, $GeCo_2O_4$ ($Co^{2+}$: $3d^7$, $S$ = 3/2) undergoes a paramagnetic-antiferromagnetic phase transition at $T_N$ = 21 K, which is accompanied with a small cubic-tetragonal structural distortion [14-17]. On the other hand, $GeNi_2O_4$ ($Ni^{2+}$: $3d^8$, $S$ = 1) shows two successive antiferromagnetic (AFM) transitions at $T_{N1}$ = 12.1 K and $T_{N2}$ = 11.4 K upon cooling, and intriguingly, neither neutron nor synchrotron x-ray diffraction measurements reveal any structural distortion at the onset of these magnetic phase transitions [14-16, 18, 19]. No structural phase transition has been observed in $GeFe_2O_4$ ($Fe^{2+}$: $3d^6$, $S$ = 2) either, although the nature of its magnetic ground state remains enigmatic [16, 20, 21]. $GeCu_2O_4$ ($Cu^{2+}$: $3d^9$, $S$ = 1/2) has a tetragonal crystal structure with drastically distorted $CuO_6$ octahedra characteristic of an elongated $c$ axis due to the strong Jahn-Teller effect of $Cu^{2+}$ ion [22], as shown in Fig. 1. $GeCu_2O_4$ undergoes an antiferromagnetic phase transition at $T_N$ ~ 33 K without any further structural distortion [22]. Based on a fit using molecular-field theory to the measured magnetic susceptibility, it was proposed that the spin structure of this material can be regarded as quasi-one-dimensional spin chains with nearest-neighboring spins aligned antiparallel to each other while the interchain coupling between neighboring $ab$ planes is much weaker [22]. However, a recent theoretical study proposed a spiral ground state based on a quasi-two-dimensional model (interacting 1-d chains) with sizable interchain coupling in the $ab$ plane, comparable with the intrachain interaction. They further suggested this compound as a possible magnetoelectric candidate [23]. Therefore, it is important to pin down the spin structure of $GeCu_2O_4$ precisely and understand how the magnetic structure is related to the distorted crystal structure.

In this article we report the magnetic properties and magnetic structure of $GeCu_2O_4$. We show that $GeCu_2O_4$ undergoes an antiferromagnetic phase transition around 32.2 K with a propagation vector of (½ ½ ½). Due to the strong Jahn-Teller distortion of $CuO_6$ octahedra and the removal of *d*-orbital degeneracy, the resulting magnetic structure consists of slabs (along the *c*-axis) composed of a pair of layers with orthogonally oriented chains; the spins of the two layers in a slab are collinear, and the spin directions of neighboring slabs are perpendicular to each other. Furthermore, we have found that the $Cu^{2+}$ spins along each chain form an unusual up-up-down-down (UUDD) configuration, which is in disagreement with the early proposals of either spiral or up-down-up-down (UDUD) spin structures [22, 23]. We propose a non-negligible nearest-neighbor biquadratic exchange interaction along the chain to explain the stability of such UUDD spin chains.

Polycrystalline $GeCu_2O_4$ studied in this work was prepared under high-pressure and high-temperature conditions using a Kawai-type multi-anvil module (Max Voggenreiter GmbH) in the Institute of Physics, Chinese Academy of Sciences. The starting materials were high-purity $GeO_2$ (99.9%) and CuO (99.9%) powders mixed in a stoichiometric ratio. These precursors were contained in a Pt capsule, inserted into an h-BN sleeve, and then placed in a graphite furnace. The whole sample assembly was put in the central hole of an MgO ceramic octahedron that was used as the pressure transmitting medium. The pressure was first increased up to 4 GPa, and then the temperature was raised to 900 ℃ and kept for 30 min, similar to the conditions reported in Ref. [22]. The temperature was then quenched to room temperature before releasing pressure slowly. The sample was recovered at ambient conditions and then subjected to various characterizations of the crystal structure and measurements of physical properties. The sample quality was checked using x-ray powder diffraction at room temperature. Heat capacity measurements were carried out using the adiabatic thermal relaxation technique on the Quantum Design PPMS cryostat. The ACMS option of the PPMS cryostat was utilized to characterize the magnetic susceptibility of $GeCu_2O_4$. Neutron powder diffraction experiments were performed using the HB-2A powder diffractometer with a wavelength of 2.41 Å and the HB-2C wide-angle neutron diffractometer with a wavelength of 1.5 Å in the High Flux Isotope Reactor (HFIR) at Oak Ridge National Laboratory.

The main panel of Fig. 2(a) shows the temperature dependence of magnetic susceptibility measured with various magnetic fields. There are three features worth pointing out. First, there is a broad peak occurring around 80 K, which is nearly independent of the magnetic field, implying the development of low dimensional short-range magnetic correlations which couple very weakly to a uniform field. Second, there is a small kink at $T_N \sim 32.2$ K in the magnetic susceptibility, as more clearly visible in the expanded view of the low field data shown in the inset of Fig. 2(b). The magnetic susceptibility measurements were conducted both after zero field cooling and field cooling procedures, nevertheless, no distinctly different behavior was observed, suggesting the absence of spin-glass like magnetic order. Instead, as to be discussed later, such a small change in magnetic susceptibility at $T_N$ stems from the onset of three-dimensional long-range magnetic ordering. In addition, one can see that even upon applying a large magnetic field up to 9 T, the transition temperature hardly changes (inset of Fig. 2(a)). Third, below $T_N$, there is an up-turn in magnetic susceptibility, which increases with an increase of magnetic field, presumably arising from a small amount of paramagnetic impurity in the material measured. Overall, these features are similar to what have been reported in Ref. [22], affirming the good quality of our sample. Figure 2(b) shows the temperature dependence of heat capacity measured at zero and 9 T. A sharp peak in heat capacity emerges at $T_N \sim 32.2$ K, indicating the occurrence of a phase transition to a long-range magnetically ordered state.

To unravel the nature of the magnetic ordering, we have performed neutron powder diffraction (NPD) measurements. Figure 3(a) presents the NPD pattern and the profile fitting of $GeCu_2O_4$ measured above $T_N$ ($T = 50$ K) on the HB2A diffractometer. Rietveld refinement of the NPD data using the FullProf Suite [24] shows that $GeCu_2O_4$ has a tetragonally distorted crystal structure with a space group of $I4_1/amd$ (Fig. 1(a)). The obtained lattice parameters, $a = b = 5.59409(4)$ Å, and $c = 9.3620$ (1) Å, are in agreement with the values reported previously [22]. Note that $c$ is much larger than $\sqrt{2}a$, indicating a much stronger structural distortion in $GeCu_2O_4$ compared with the cubic-tetragonal structural distortion simultaneously taking place at the onset magnetic ordering observed in various spinel compounds [1]. The strong tetragonal distortion arises from the significant Jahn-Teller distortion associated with $Cu^{2+}$ ions, which results in elongated $CuO_6$ octahedra along the $c$-axis (Fig. 1(b)). Consequently, the Cu-Cu bond lengths of each Cu tetrahedron breaks into two types as shown in Fig. 1(c), one type is shorter with bond length ($\sim 2.797$ Å) in the $ab$ plane compared to the other along the $c$-axis ($\sim 3.064$ Å). This results in a weaker coupling between Cu chains lying in different $ab$ planes.

The NPD pattern measured below $T_N$ ($T = 2$ K) and the Rietveld fit profile are presented in Fig. 3(b). An expanded view of the low $Q$ region is shown in the inset. One can clearly see several magnetic Bragg peaks showing up at low temperature, which can be well indexed with a magnetic propagation vector $\mathbf{k}$ = (½ ½ ½) based on the nuclear structure and indicate antiferromagnetic characteristic of the magnetic ground state. Figure 4 displays the temperature dependence of the order parameter obtained from the (½ ½ ½) Bragg peak intensity measured on the HB2C diffractometer. Fitting the results to a power-law function (red curve) gives a transition temperature of 32.9 K, which is close to the value determined from heat capacity and magnetic susceptibility measurements.

Possible models of the magnetic structure have been explored by representation analysis using the program *BasIreps* in the FullProf Suite [24], and by magnetic symmetry approach using the MAXMAGN available at the Bilbao Crystallographic Server [25]. The schematics of the most appropriate spin structure model that describes the NPD pattern is illustrated in Fig. 5(a). The corresponding fitting profile is shown by the red curves in Fig. 3(b) and its inset. The goodness of the fit $R_{wp}$ is 12.2. The proposed spin arrangement associated with the 2a, 2b, 2c-cell enlargement (i.e. **k** = (½, ½, ½)) can be described by the magnetic space group $I_c$-42$d$. The corresponding magnetic configuration for each atom position is summarized in Table 1. The refined magnetic momentum is ~ 0.89(5) $\mu_B$ per Cu, which is slightly smaller than the theoretical value of 1 $\mu_B$ per Cu for fully localized electronic state of $Cu^{2+}$ ions with $S = ½$. The reduction in the moment can be due to the covalency or quantum fluctuations enhanced by magnetic frustration and the low dimensionality. The magnetic moments are lying in the basal plane, with the out of the plane component refined to be within the fit uncertainty, less than 0.05 $\mu_B$ / Cu.

The magnetic structure shown in Fig. 5 exhibits several important features. First, the system displays chain-like structure along both *a* and *b* directions in the *ab* plane (Fig. 5(b)). This mainly originates from the Jahn-Teller distortion of the $Cu^{2+}$ ions, which leads to a shorter nearest-neighbor (NN) Cu-Cu distance along the chains in the *ab* plane and a longer Cu-Cu distance along the out-of-plane direction as discussed above. Also due to the spinel structure the distance between chains in a given *ab* plane is twice of the shortest Cu-Cu bond length. Since the one hole of $Cu^{2+}$ ions mostly occupies the $d_{x^2-y^2}$ orbital, there are strong magnetic interactions between Cu ions along the in-plane directions and weaker interlayer coupling. Second, intriguingly, spins along each chain form an unusual UUDD pattern (Fig. 5(b)), which is in sharp contrast to an early prediction of either a spiral spin structure [23] or UDUD collinear ordering [22]. Possible underlying mechanism for the formation of the UUDD spin structure along the chain will be discussed in the next paragraph. Third, as illustrated in Fig. 5(b), the interchain spins within the *ab* plane are aligned antiparallel, implying an antiferromagnetic coupling via the Cu-O-O-Cu superexchange path, which is consistent with a recent density functional theory (DFT) calculation that reported this interaction value to be 130 K [23]. Fourth, as shown in Fig. 5(c), the interlayer spins within the tetrahedron labeled as circle 1 (two layers in a slab) are aligned antiparallel to each other, indicating an antiferromagnetic interlayer coupling which is not consistent with the predicted ferromagnetic interlayer coupling [23]. Nevertheless, interestingly, the interlayer spins within the tetrahedron label as circle 2 in Fig. 5(c) align perpendicularly to each other, which will be further discussed later.

The existence of a UUDD phase has been found in an exact solution for the ground state of a frustrated *classical* Heisenberg chain with ferromagnetic NN interaction, antiferromagnetic next-nearest-neighbor (NNN) interaction, and a NN biquadratic exchange interaction [26]. However, the formation of a UUDD magnetic structure for spin ½ $Cu^{2+}$ chains is unusual. Recent DFT calculations for $GeCu_2O_4$ have shown the existence of magnetic frustration along the chain with a ferromagnetic NN coupling of ~ -60 K and an antiferromagnetic NNN coupling of ~ 80 K, which accounts for the predicted spiral spin structure [23]. The ferromagnetic NN coupling arises from a combination of the direct exchange interaction between $Cu^{2+}$ ions and the superexchange interaction via Cu-O-Cu with a bond angle of 91.21º. The discrepancy between the predicted spiral spin structure and the observed UUDD spin order along the chain suggests that one needs to take into account additional sizeable exchange interactions other than ferromagnetic NN and antiferromagnetic NNN couplings. One very likely candidate is the biquadratic NN exchange, the importance of which has been previously discussed in classical spin systems [26, 27]. Rigorously, for a quantum spin $S = ½$ the biquadratic exchange term $(\vec{S}_i \cdot \vec{S}_{i+1})^2$ can be rewritten as $A + B(\vec{S}_i \cdot \vec{S}_{i+1})$ where A and B are constants, reducing the biquadratic exchange to a bilinear exchange, which would not affect the magnetic structure obtained in the bilinear isotropic Heisenberg Hamiltonian. Nevertheless, such a scenario may not be true when the spin-orbit coupling (SOC) is not negligible. Recently, there has been a growing interest in investigating the effect of SOC in $Cu^{2+}$ systems using DFT calculations. SOC gives rise to either the anisotropic superexchange interaction of the nearly rectangular Cu-O-Cu bond [28, 29] or single-ion anisotropy [30,31]. Due to the Jahn-Teller distortion in $GeCu_2O_4$, there is a singly occupied nondegenerate $d_{x^2-y^2}$ orbital, which entangles with a doubly occupied $d_{xy}$ orbital via SOC. This then gives 4 possible states for the one hole (or 3 electrons) per $Cu^{2+}$ ion; thus, one can describe the system with a pseudospin $S = 3/2$ and this generalized spin can support a biquadratic term in the effective spin Hamiltonian. Being $S = 3/2$ one can expect a nonzero nearest neighbor biquadratic exchange which can lead to the observed UUDD spin structure along the chain. Detailed calculations are necessary to check this speculation and estimate the strength of this exchange term.

Finally, we would like to briefly comment on the perpendicular spin alignment of the interlayer spins within the tetrahedron 2 illustrated in Fig. 5(c). One the one hand, one can see that tetrahedron 1 and tetrahedron 2 corner-share a $Cu^{2+}$ ion while spins along the chain between two NN tetrahedra are antiferromagnetically aligned (due to the UUDD pattern discussed above). This gives rises to a magnetic frustration between the interlayer spins within tetrahedron 2. On the other hand, the SOC discussed above leads to the observed magnetic anisotropy in this system with the spins pointing along the diagonal direction in the *ab* plane. These two factors, combined with the Dzyaloshinskii-Moriya (DM) interaction associated with the local inversion symmetry breaking of the Cu-Cu bond along the *c*-axis, can account for the perpendicular spin alignment. This further implies that the DM interaction is comparable in strength with the interlayer coupling, both of which are much weaker than the intralayer interactions, thus having no effect on the in-plane collinear spin structure. Calculations using perturbation theory with generalized Hilbert space or DFT calculations with SOC and allowing for the UUDD state are warranted to further understand the underlying physical mechanisms which drive the magnetic structure in this Jahn-Teller distorted pyrochlore $S = ½$ system.

In summary, we have shown that the magnetic structure of a tetragonally distorted spinel $GeCu_2O_4$ can be described as slabs composed of a pair of layers with orthogonally aligned chains and collinear spin configurations between the two layers, while the spin direction of neighboring slabs along the *c*-axis are perpendicular to each other. The observed UUDD spin structure along each chain suggests a non-negligible biquadratic exchange interaction in the effective spin Hamiltonian. This work calls for theoretical studies which investigate carefully the effect of spin-orbit coupling and inter-orbital mixing of $Cu^{2+}$ ions in understanding the magnetic structure of this unusual and exciting material.

X. K. acknowledges the support from the start-up funds at Michigan State University and appreciates insightful discussion with Prof. G.-W. Chern. J.G.C. acknowledges the support of the NNSF and MOST of China (Grant Nos. 11304371, 11574377, and 2014CB921500), the Strategic Priority Research Program and the Key Research Program of Frontier Sciences of the Chinese Academy of Sciences (Grant No. XDB07020100), and the Opening Project of Wuhan National High Magnetic Field Center (Grant No. 2015KF22), Huazhong University of Science and Technology. Work at ORNL was supported by the Scientific User Facilities Division, Office of Basic Energy Sciences, DOE.

TABLE 1: Arrangement of the magnetic moments for the Cu atoms occupying the 8c (0,0,0) position of the parent space group of $I4_1/amd$. The actual magnetic unit cell is defined by the propagation vector **k** = (½, ½, ½).

| Atom # | Atomic coordinates | Relative moments projections |
|---|---|---|
| 1 | 0, 0, 0, | mx, my, mz |
| 2 | 1/2, 0, 1/2 | -mx, my, -mz |
| 3 | 3/4, 1/4, 3/4 | -my, mx, -mz |
| 4 | 3/4, 3/4, 1/4 | -my, -mx, -mz |
| 5 | 1/2, 1/2, 1/2 | -mx, my, mz |
| 6 | 0, 1/2, 0 | -mx, -my, mz |
| 7 | 1/4, 3/4, 1/4 | my, mx, -mz |
| 8 | 1/4, 1/4, 3/4 | -my, mx, mz |

Figure captions:

Figure 1. (a) Schematics of the crystal structure of GeCu$_2$O$_4$. (b) Expanded view of CuO$_6$ octahedra showing the bond length and bond angle. (c) Schematic of Cu tetrahedra with two different Cu-Cu bond length of each tetrahedron.

Figure 2. (a) Temperature dependence of magnetic susceptibility measured with various magnetic fields, inset shows an expanded view of low temperature regime. (b) Temperature dependence of specific heat measured at zero and 9 T magnetic field. Inset shows an expanded view of magnetic susceptibility measured with 5000 Oe magnetic field.

Figure 3. (a) Neutron powder diffraction pattern (black dot) and Rietveld refinement (red curve) of GeCu$_2$O$_4$ measured at 50 K. The Bragg peaks of aluminum sample holder are also refined as indicated by the second row of Bragg positions (olive vertical lines). (b) Neutron powder diffraction pattern and Rietveld refinement results at 2 K. The magnetic Bragg peak positions are indicated by the third row of Bragg peak positions (olive vertical lines). The inset shows the low-Q region magnetic Bragg peaks obtained from the difference $I_{\text{diff}} = I_{2K} - I_{50K}$ and its Rietveld refinement.

Figure 4. Peak intensity of $Q$ = (½ ½ ½) magnetic Bragg peak as a function of temperature and the fitting to a power law.

Figure 5. (a) Refined magnetic structure of GeCu$_2$O$_4$ within one magnetic unit cell. (b) *ab* plane projection of the spin structure showing the up-up-down-down configuration along the chain. (c) Zoom-in view of the spin structure of three corner-sharing Cu tetrahedra.

Figure 1

T. Zou et al

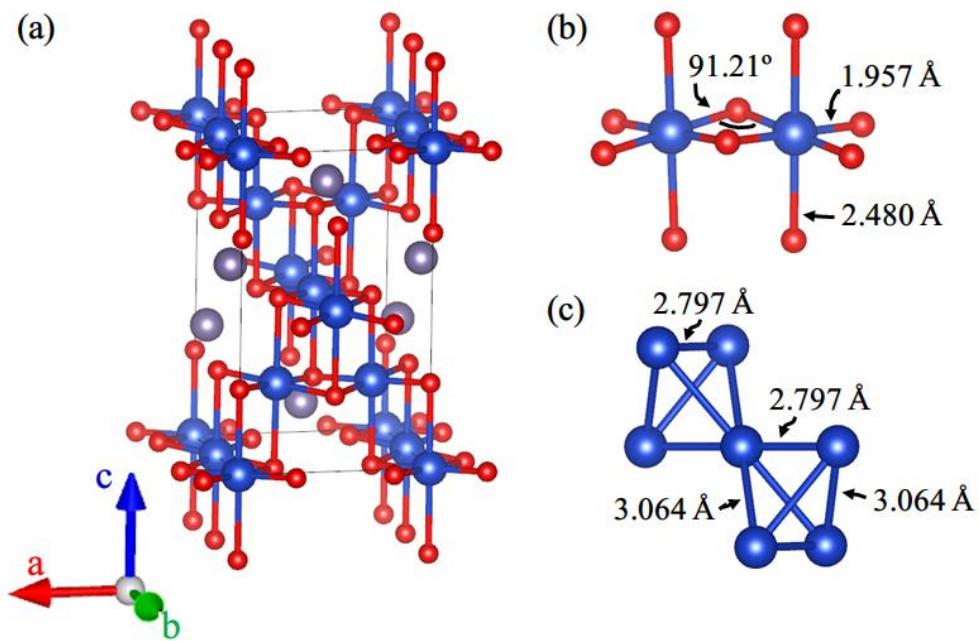

Figure 2

T. Zou et al

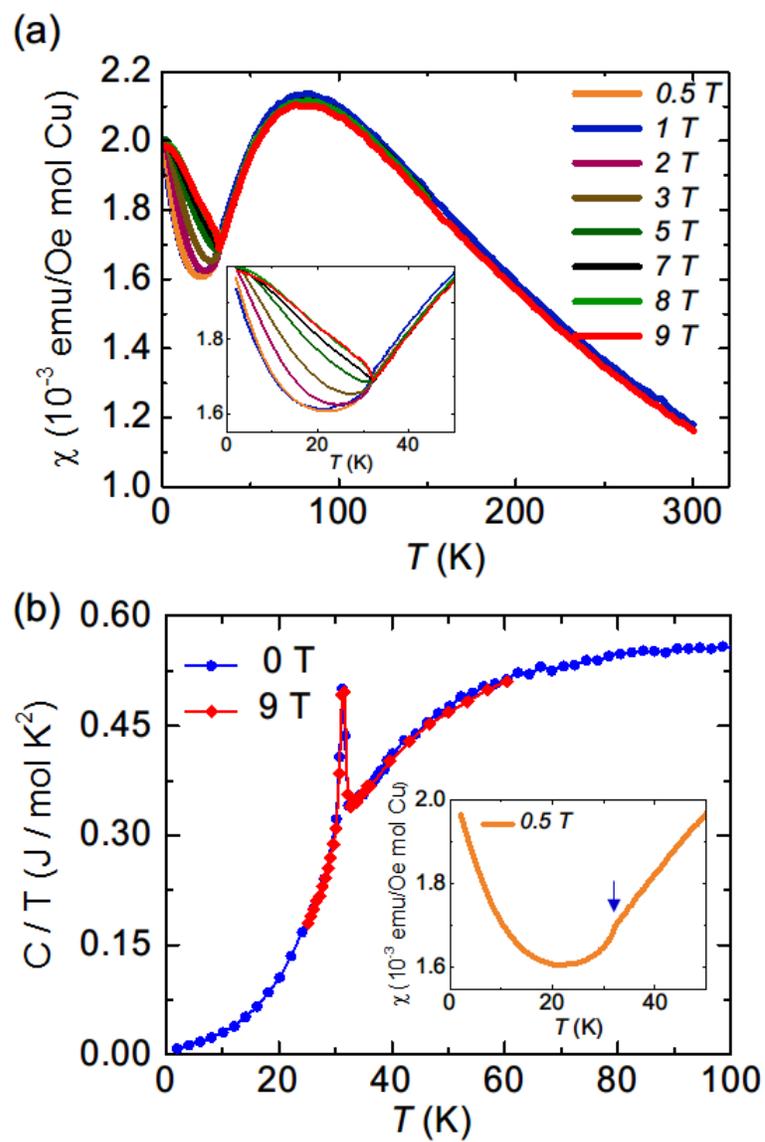

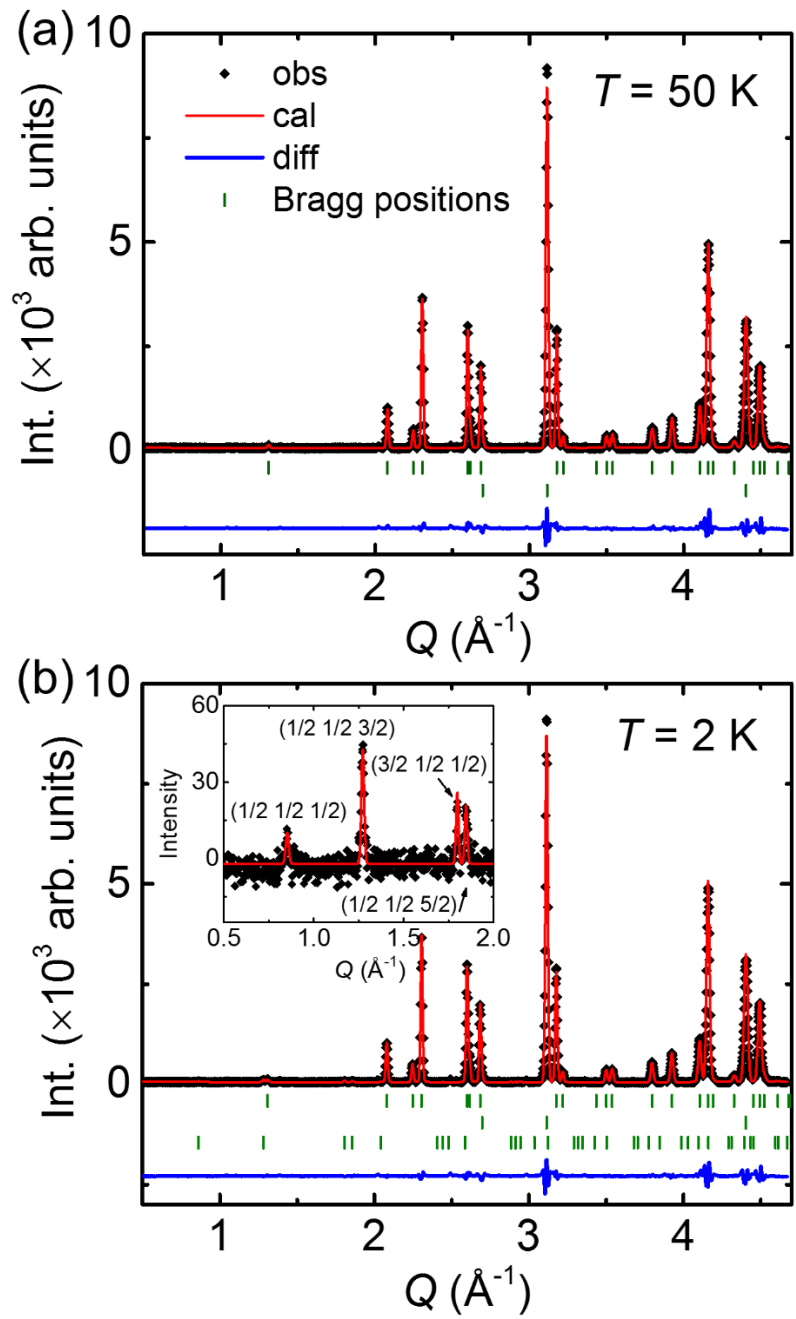

Figure 4

T. Zou et al

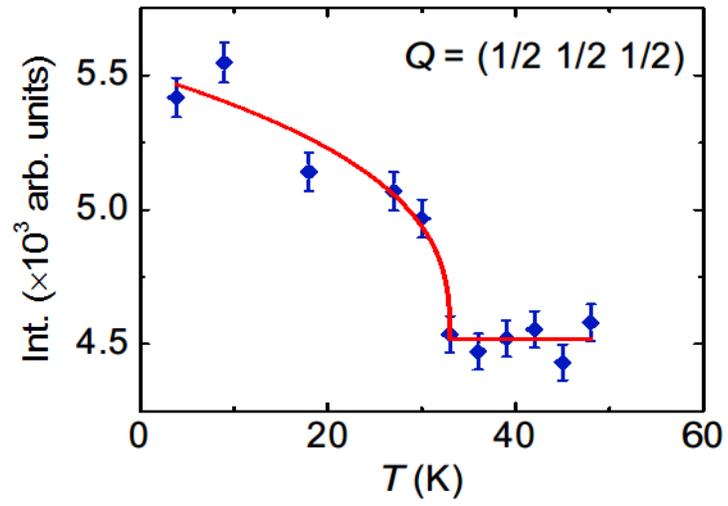

Figure 5

T. Zou et al

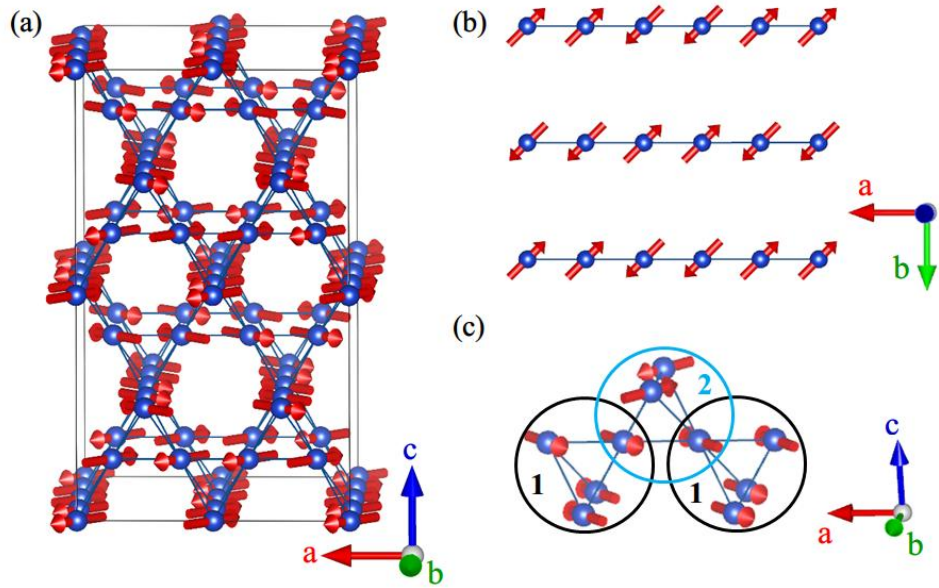

[31] Purely spin ½ ions such as $Cu^{2+}$ where there is one electron (or hole) occupying a nondegenerate orbital, cannot have a single ion anisotropy in their spin Hamiltonian because $S_x^2 = S_y^2 = S_z^2 = 1/4$. However DFT calculations for several $Cu^{2+}$ systems including spin-orbit coupling by J. Liu et al [Ref. 30] gives a finite single ion anisotropy. This can be reconciled by noting that in DFT calculations (essentially a mean-field theory) the average of spin density is the physical variable which can be treated as a classical in which case a single ion anisotropy term is possible.